\documentclass[12pt]{iopart}
\pdfoutput=1
\usepackage{graphicx}
\usepackage{cite}
\usepackage{amsfonts}
\usepackage{wrapfig}
\usepackage{srcltx}
\usepackage{color}

\begin{document}

\title[Entropy Production of Irreversible Transitions]
      {Entropy Production of Nonequilibrium Steady States with Irreversible Transitions}
\author{Somayeh Zeraati$^1$, Farhad H. Jafarpour$^1$, and Haye Hinrichsen$^2$}
\address{$^1$Bu-Ali Sina University, Physics Department, 65174-4161 Hamedan, Iran}
\address{$^2$Universit\"at W\"urzburg, Fakult\"at f\"ur  Physik und Astronomie, 97074  W\"urzburg, Germany.}

\begin{abstract}
In nature stationary nonequilibrium systems cannot exist on their own, rather they need to be driven from outside in order to keep them away from equilibrium. While the internal mean entropy of such stationary systems is constant, the external drive will on average increase the entropy in the environment. This external entropy production is usually quantified by a simple formula, stating that each microscopic transition of the system between two configurations $c \to c'$ with rate $w_{c\to c'}$ changes the entropy in the environment by $\Delta S_{\rm env} = {\ln w_{c \to c'}}-{\ln w_{c' \to c}}$. According to this formula  irreversible transitions $c \to c'$ with a vanishing backward rate $w_{c'\to c}=0$ would produce an infinite amount of entropy. However, in experiments designed to mimic such processes, a divergent entropy production, that would cause an infinite increase of heat in the environment, is not seen. The reason is that in an experimental realization the backward process can be suppressed but its rate always remains slightly positive, resulting in a finite entropy production. The paper discusses how this entropy production can be estimated and specifies a lower bound depending on the observation time.
\end{abstract}

\def\d{{\rm d}}
\def\0{\emptyset}
\def\mark#1{{\color{red}#1\color{black}}}

\ead{farhad@ipm.ir}

\parskip 2mm 

%==========================================================================
\section{Introduction}
%==========================================================================

In classical statistical physics, complex systems are often modeled as continuous-time Markov processes in a discrete configuration space. Such systems evolve by spontaneous random transitions from configuration $c$ to configuration $c'$ according to certain transition rates $w_{c\to c'}$. For isolated systems these rates are symmetric so that the model evolves into an equilibrium state with maximal entropy. For open systems, which interact with the surrounding environment, the transition rates are generally asymmetric. In this case the dynamical evolution changes not only the system's entropy but also the entropy in the environment. The entropy change in the environment is referred to as the \textit{entropy production} of the system.

As shown by Schnakenberg, Andrieux, Gaspard and Seifert~\cite{Schnakenberg,Gaspard,Seifert,Seifert2}, the entropy in the environment $S_{\rm env}$ changes discontinuously by
\begin{equation}
\label{EntropyProduction}
 \Delta S_{\rm env} \;=\; \ln \frac{w_{c\to c'}}{w_{c'\to c}}\,
\end{equation}
whenever the system jumps from $c$ to $c'$ (note that we set $k_B=1$ for simplicity). This simple formula is independent of the specific composition and structure of the environment, provided that it equilibrates almost instantaneously between successive transitions of the Markov process~\cite{Kolkata,SM}.

The entropy production formula (\ref{EntropyProduction}) requires that for any transition $c \to c'$  with a non-vanishing forward rate $w_{c \to c'}>0$, the corresponding backward rate $w_{c' \to c}$ has to be nonzero as well since otherwise the entropy production would diverge. This means that the definition of entropy production is only meaningful in models with microscopically reversible transitions. Usually it is argued that in realistic physical systems the effective backward rate is always nonzero since the classical description results from a coarse-graining of the underlying quantum-mechanical processes which are intrinsically time-reversible.

However, in statistical physics a large variety of models investigated in the literature involve microscopically irreversible transitions. For some of these models experiments have been suggested or performed. A simple example is the totally asymmetric simple exclusion process, where particles hop only in one direction, which is studied experimentally by optical tweezers~\cite{Korda,Lacasta,Dickman}. Another example is directed percolation~\cite{Haye}, the standard model of a phase transition from an fluctuating phase into a frozen state, which was recently realized experimentally for the first time~\cite{Takeuchi}. However, a divergent entropy production, which would manifest itself in form of a divergent increase of heat in the environment, has not been reported in these experiments. This suggests that this divergence is a theoretical artifact and needs to be regularized in a meaningful way.

To our knowledge ben-Avraham, Dorosz and Pleimling~\cite{benAvraham} were the first to address this problem in detail. As a possible solution they suggest to coarse-grain the stochastic evolution by interval sampling: Instead of monitoring each transition event separately, they propose to read off the configuration at regular temporal intervals and to use the resulting configuration sequence to define effective transition rates. Even if the backward rate $w_{c'\to c}$ is zero, meaning that direct transitions from $c'$ to $c$ are forbidden, the sampling allows the system to evolve from $c'$ to $c$ through a loop of other intermediate configurations between two consecutive readings. This gives rise to a small but finite effective backward rate in the sampled data, regularizing the entropy production depending on the sampling rate.

In this Letter we propose an alternative regularization method which is more closely related to the question how micro-irreversibility can be implemented in experiments. We start with the assertion that Eq.~(\ref{EntropyProduction}) is indeed correct and that the preceding argument about the fundamental impossibility of vanishing backward rates in nature remains valid. This means that it is \textit{in principle} impossible to realize micro-irreversible processes experimentally. However, in practice one can approximate irreversible processes very well by designing the experiment in such a way that the backward transition is strongly suppressed. In such an experiment the actual backward rate is positive but so small that the reverse transition practically never takes place during data taking. Nevertheless the positivity of this rate ensures that the entropy produced by the corresponding forward process is still finite.

The aim of this work is to specify a lower bound on the entropy production of a physical system which is designed to approximate an irreversible process over a finite time span~$T$. We find that the entropy production rate of such a system can be split into two parts, namely, a conventional constant part stemming from the reversible transitions, and a second part coming from the approximated irreversible transitions which grows logarithmically with~$T$. 

%==========================================================================
\section{Lower bound on the entropy production of micro-irreversible transitions}
%==========================================================================

Starting point is a continuous-time Markov process defined by certain set of configurations $c \in \Omega$. The model evolves by random transitions $c \to c'$ with rates $w_{c \to c'} \geq 0$, where some of the allowed transitions are microscopically irreversible, i.e. $w_{c' \to c}=0$. Suppose that we are able to design an experiment with an identical set of possible configurations which approximately reproduces the Markovian dynamics of the model. In what follows let us distinguish between
\begin{quote}
\begin{itemize}
 \item the \textit{defining} rates  $w_{c\to c'}$ of the original model, and
 \item the corresponding \textit{actual} rates $\tilde w_{c\to c'}$ realized in the experiment.
\end{itemize} 
\end{quote}
However, usually the actual rates in the experiment are not directly accessible, rather they have to be estimated from the observed number of transitions $n_{c \to c'}$ during a finite time span $T$ of data taking. If the system is found in the configuration $c'$ with probability~$P_{c'}$, the expectation value of this number is given by $P_{c'} \tilde w_{c'\to c} T$. Observing a vanishing number $n_{c' \to c}=0$ in a single experiment does not necessarily imply that the corresponding actual rate $\tilde w_{c' \to c}$ is zero, it only means that this rate is sufficiently smaller 
than $(P_{c'}T)^{-1}$ so that this transition did not occur during data taking.

In the following we use this framework to specify a lower bound for the entropy production caused by irreversible transitions in experiments with a finite observation time. The idea is to estimate the actual rate $\tilde w$ of a transition $c \to c'$ in the experiment for a given defining rate $w$ on the basis of the expected count numbers $n_{c \to c'}$ within a given observation time $T$. In this way we want to find a physically motivated conditional probability distribution $P(\tilde w|w)$ of the actual rate $\tilde w$ for a given defining rate $w$.

To determine $P(\tilde w|w)$ let us assume that the transition $c \to c'$ occurs $n$ times during the observation time $T$. As these events are spontaneous and uncorrelated, $n$ is randomly distributed according to a Poisson distribution
\begin{equation}
\label{Poisson}
P(n|w)=\frac{(\tau w)^ne^{-\tau w}}{n!}\,,
\end{equation}
where $\tau = P_{c}T$ is the expected time that the system spends in the configuration $c$. This allows us to express $P(\tilde w|w)$ as
\begin{equation}
P(\tilde w|w) = \sum_{n=0}^\infty P(\tilde w|n)P(n|w)\,,
\end{equation}
where $P(\tilde w|n)$ is the likelihood for the distribution of the actual rate $\tilde w$ for a given number of transitions $n$. According to Bayes rule~\cite{Bayes} this likelihood is given by
\begin{equation}
P(\tilde w|n)=\frac{P(n|\tilde w)P(\tilde w)}{P(n)} \,,
\end{equation}
where $P(\tilde w)$ is the \textit{prior distribution} and 
\begin{equation}
\label{prior}
P(n)=\int_0^\infty\d \tilde w\,P(n|\tilde w)P(\tilde w)
\end{equation}
the normalizing marginal likelihood. The prior $P(\tilde w)$ expresses our belief of how the rates are typically distributed and therefore introduces a certain degree of ambiguity in the derivation. However, if no specific information about this distribution is available, it is customary to use the so-called \textit{conjugate prior} which ensures that the posterior $P(\tilde w|n)$ and the prior $P(\tilde w)$ belong to the same family of distributions. The conjugate prior of a Poisson likelihood distribution $P(n|\tilde w)$ is the Gamma distribution
\begin{equation}
\label{Prior}
P(\tilde w)=\frac{\tilde\beta^{\tilde\alpha} \tilde w^{\tilde\alpha-1} e^{-{\tilde\beta} \tilde w}}{\Gamma(\tilde\alpha)}
\end{equation}
which depends on two hyperparameters $\tilde\alpha$ and $\tilde\beta$ (the tilde is used to avoid confusion with rates $\alpha,\beta$ for various models used in the literature). With this prior the posterior is given by
\begin{equation}
P(\tilde w|n) \;=\; \frac{\tilde w^{\tilde\alpha +n-1} e^{-(\tilde\beta+\tau)  \tilde w} (\tilde\beta +\tau)^{\tilde\alpha +n}}{\Gamma
   (n+\tilde\alpha )}
\end{equation}
which allows one to compute the expectation value of the actual rate for the transition $c \to c'$
\begin{equation}
\label{Average}
\langle \tilde w \rangle = \int \d \tilde w \, \tilde w P(\tilde w|w)= \frac{\tau w+\tilde\alpha}{\tau +\tilde\beta}\,.
\end{equation}
As can be seen, this formula maps a vanishing defining rate $w=0$ onto a nonvanishing actual rate $\langle \tilde w \rangle \propto 1/\tau$.
By defining $\tau'=P_{c'} T$ and inserting this expectation value into the entropy production formula~(\ref{EntropyProduction}) and taking the 
limit $T \gg 1$ the entropy production for reversible transitions
\begin{eqnarray}
\label{EstimateEPR}
\Delta S_{\rm env}^{\rm rev}(c\to c') & = & \ln \frac{\langle \tilde w_{c\to c'} \rangle}{\langle \tilde w_{c'\to c} \rangle}  \nonumber \\
& = &  \ln \frac{(\tau w_{c \to c'}+\tilde\alpha)/(\tau +\tilde\beta)}{(\tau' w_{c' \to c}+\tilde\alpha)/(\tau' +\tilde\beta)} \;\approx\; \ln \frac{w_{c\to c'}}{w_{c'\to c}}
\end{eqnarray}
reproduces the known result in Eq.~(\ref{EntropyProduction}). As our main result, for irreversible transitions we obtain a finite entropy production which grows logarithmically with the observation time
\begin{eqnarray}
\label{EstimateEPIR}
\Delta S_{\rm env}^{\rm irr}(c\to c') & = &  \ln \frac{\langle \tilde w_{c\to c'} \rangle}{\tilde\alpha/(\tau+\tilde\beta)} \nonumber \\
& = & \ln \frac{(\tau w_{c \to c'}+\tilde\alpha)/(\tau +\tilde\beta)}{\tilde\alpha/(\tau' +\tilde\beta)} 
\;\approx\; \ln \frac{\tau' w_{c\to c'}}{\tilde\alpha} \,.
\end{eqnarray}
Note that we have simplified the derivation by replacing $\langle \ln \tilde w\rangle \to \ln \langle \tilde w \rangle$. However, as shown in Sect. 3 of the Supplemental Material~\cite{SM}, apart from a redefinition of $\tilde\alpha$, a correct derivation to lowest order leads to the same result.

The prior distribution (\ref{Prior}) depends on two hyperparameters $\tilde\alpha$ and $\tilde\beta$ which determine its shape and scale. Since the entropy production depends on a ratio of rates, the scale hyperparameter $\tilde\beta$ drops out. However, the shape hyperparameter $\tilde\alpha$ appears in the final result and thus it has to be defined in a physically meaningful way. In this regard note that the Gamma distribution~(\ref{Prior}) evaluated at the origin is finite for $\tilde\alpha=1$, infinite for $\tilde\alpha<1$ and zero for $\tilde\alpha>1$. Therefore, in experiments of models with irreversible transitions, where the likelihood of a vanishing defining rate is expected to be finite, the most natural choice, which we will use from now on, is $\tilde\alpha=1$.

%==========================================================================
\section{Examples}
%==========================================================================

In what follows we study the entropy production in three exemplary nonequilibrium systems with micro-irreversible transitions in their steady state (further examples can be found in the Supplemental Material~\cite{SM}). To this end we first determine the stationary probability distribution $P_c$ to find the system in configuration $c$. The average entropy production is then given by
\begin{equation}
\langle \dot S_{\rm env} \rangle \;=\; \sum_{c \neq c'} P_c \, w_{c \to c'} \, \Delta S_{\rm env}(c \to c')\,,
\end{equation}
where one has to sum over $\Delta S_{\rm env}^{\rm rev}(c\to c')$ or $\Delta S_{\rm env}^{\rm irr}(c\to c')$ depending on whether the transition $c \to c'$ is reversible or irreversible.  For lattice models with $L$ sites, we define the average entropy production per site
\begin{equation}
\label{AverageEntropyProduction}
\dot s_{\rm env} \;:=\; \frac{1}{L} \langle \dot S_{\rm env} \rangle\,.
\end{equation}
%
%
%-----------------------------------------------------------------
%
{\bf TASEP:} \\
The first example is the Totally Asymmetric Simple Exclusion Process (TASEP) on a finite lattice with $L$ sites, where particles are added at the left boundary (removed from the right boundary) with rate $\alpha$ ($\beta$) if the first (last) lattice site is empty (occupied). Since particles hop exclusively to the right, all microscopic transitions are irreversible. The stationary probability distribution of the TASEP can be calculated exactly using the matrix product method~\cite{DEHP}, where the stationary weight of each configuration is determined by a product of noncommuting operators corresponding to the actual configuration. Using the exact results of Ref.~\cite{DEHP} in the formulas~(\ref{AverageEntropyProduction}) and~(\ref{EstimateEPIR}) it turns out that, to leading order in $T$, the large-$L$ limit of the average entropy production rate
\begin{equation}
\label{TASEPEntropyProduction}
\dot s_{\rm env}  \;\approx\; J(\alpha,\beta) \ln T 
\end{equation}
is proportional to the particle current $J(\alpha,\beta)$ in the steady state. Depending on the values of $\alpha$ and $\beta$ the current is equal to $J(\alpha,\beta)=\frac{1}{4}$ for $\alpha,\beta \ge \frac{1}{2}$, $J(\alpha,\beta)=\alpha(1-\alpha)$ for $\alpha<\beta$ and $\alpha<\frac{1}{2}$, and $J(\alpha,\beta)=\beta(1-\beta)$ for $\beta<\alpha$ and $\beta<\frac{1}{2}$. Thus the average entropy production changes continuously in the parameter space and attains its maximum for $\alpha,\beta \ge \frac{1}{2}$ where the particle current in maximal. 

%-----------------------------------------------------------------

%======================================================
\begin{figure}[t]
\centering\includegraphics[width=100mm]{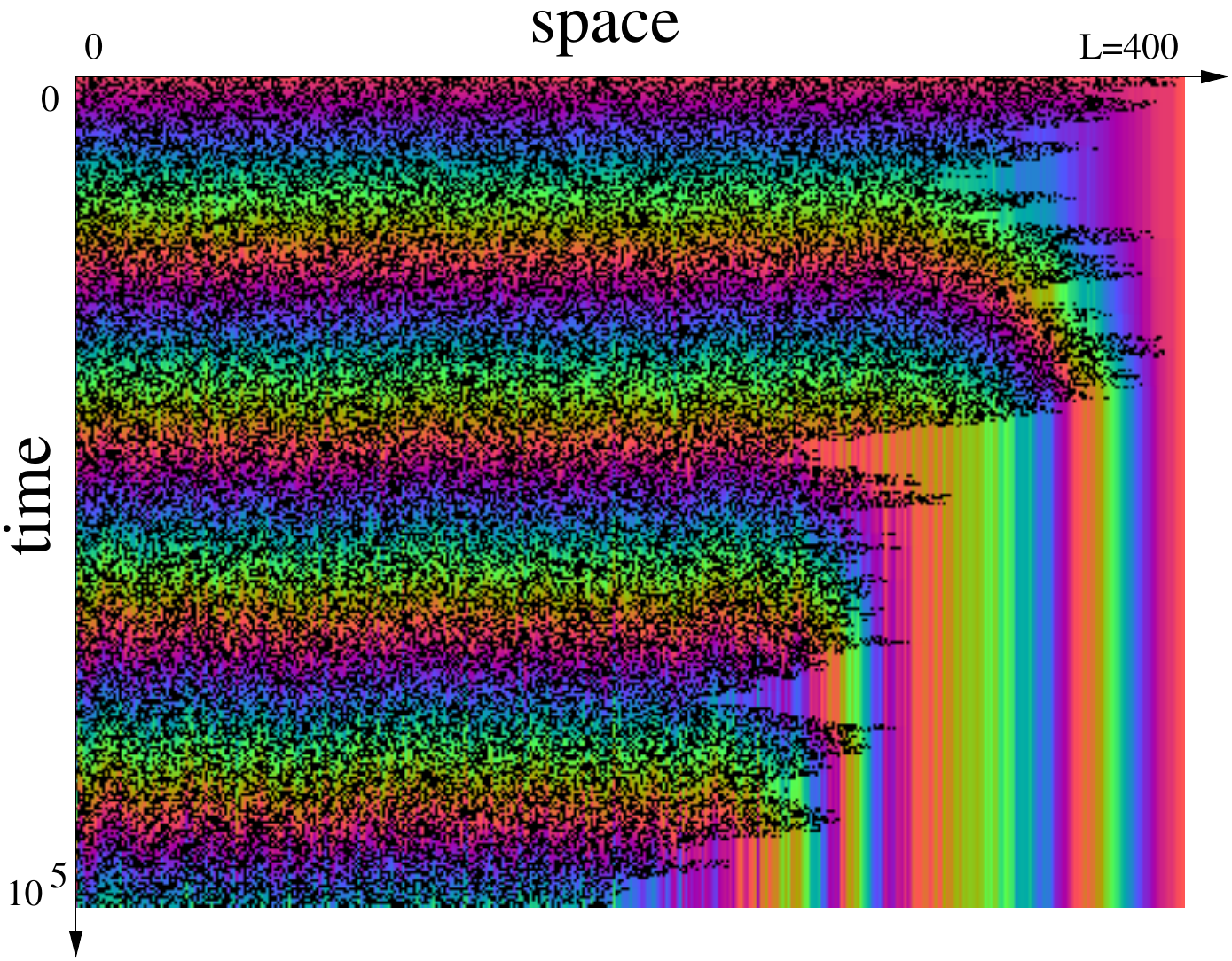}
\caption{(color online) Branching-Coalescing Process (BCP) at the critical point, exhibiting a diffusing Bernoulli shock. Particles are represented by black pixels while the integrated entropy production at each site is visualized by a periodically changing color scale. \label{fig:bcp}}
\end{figure}
%======================================================

\noindent {\bf BCP:} \\
The second example is the Branching-Coalescing Process (BCP) on a one-dimensional lattice with $L$ lattice sites and open boundaries. Each lattice site is either empty ($\0$) or occupied by at most one particle ($1$). In the bulk the BCP evolves by the dynamical rules
\begin{equation}
\label{RULESBCP}
\begin{array}{l}
\0 1\to 11\;\mbox{with rate}\;w\\
1 1 \to 1\0\;\mbox{with rate}\;w\\
11\to \0 1\;\mbox{with rate}\;1\\
1 \0\to 11\;\mbox{with rate}\;1\\
1 \0\to \01\;\mbox{with rate}\;1.
\end{array}
\end{equation}
In addition, particles are added at (removed from) the left boundary with rate $\alpha\;$ ($\gamma$) while at the right boundary particles are removed with rate $\beta$. 

It is known that the steady state of the BCP can be written as a linear superposition of Bernoulli shock measures provided that $\gamma=\alpha+\frac{w}{2}-1$~\cite{KJS}. Moreover, under the same constraint it turns out that the model has a matrix product steady state~\cite{Jafarpour}. Varying $w$ the process undergoes a phase transition between a high- and a low-density phase at $w_c=4$. Applying the matrix product method with the two-dimensional representation introduced in~\cite{Jafarpour}, it is straightforward to calculate the average entropy production rate per site in the steady state. To leading order in $T$ and in the large-$L$ limit one finds
\begin{equation}
\label{APBCP}
\dot s_{\rm env}  \;\approx\; 
\left\{
\begin{array}{ll}
\frac{1}{4}\ln T& \mbox{for} \; 2(1-\alpha) < w<w_{c}, \\[2mm] 
0                                                  & \mbox{for} \; w>w_{c}.
\end{array}
\right.
\end{equation}
As can be seen, the average entropy production of the BCP changes \textit{discontinuously} at the transition point. The nonzero part of the entropy production 
in~(\ref{APBCP}) comes from both reversible and irreversible processes in~(\ref{RULESBCP})~\cite{SM}. In Fig. \ref{fig:bcp} the time evolution of the entropy production at each lattice site is plotted at the critical point $w=w_{c}$. In this case the last particle in the system (the last black pixel from the left) performs an unbiased random walk on the lattice. It can be seen that only occupied lattice sites contribute to entropy production. 

%-----------------------------------------------------------------

%======================================================
\begin{figure}[t]
\centering\includegraphics[width=110mm]{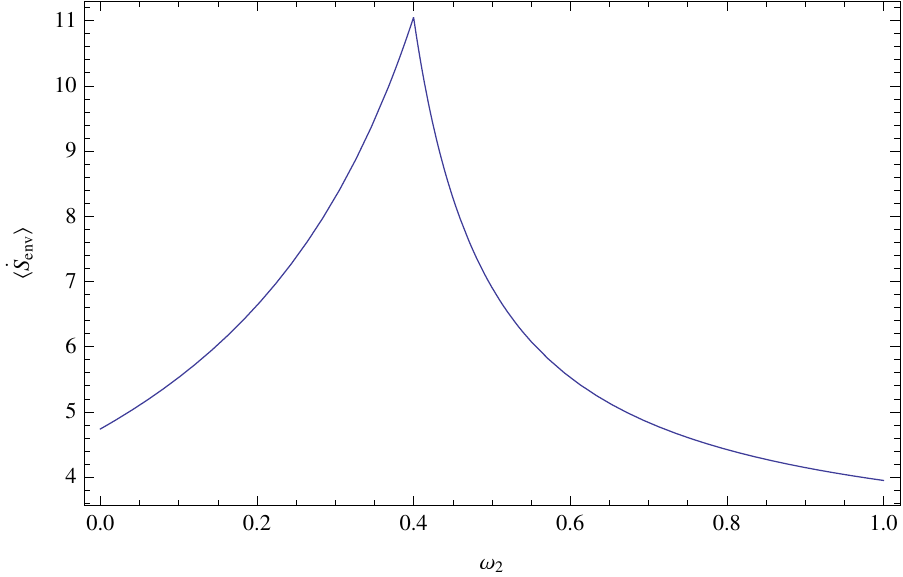}
\vspace{-3mm}
\caption{Average irreversible entropy production in the AKGP as a function of $w_{2}$ for $w_{1}=0.4$, $\alpha=0.3$, $\beta=0.1$, $L=10^{6}$, and $T=10^{6}$. The phase transition occurs at $w_{2}=0.4$.\label{fig:akgp}}
\end{figure}
%======================================================

\noindent {\bf AKGP:} \\
The third example is the one-dimensional Asymmetric Kawasaki-Glauber Process (AKGP) with the dynamical rules
\begin{equation}
\label{RULESAKGP}
\begin{array}{l}
1 \0\to\0\0\;\mbox{with rate}\;w_{1}\\
1 \0\to11\;\mbox{with rate}\;w_{2}\\
\01\to\0\0\;\mbox{with rate}\;w_{3}\\
\01\to11\;\mbox{with rate}\;w_{4}\\
\01\to1\0\;\mbox{with rate}\;w_{5}.
\end{array}
\end{equation}
In addition, particles are added at (removed from) the left (right) boundary with rate $\alpha\;(\beta)$. In this model all transitions are irreversible. It is known that the AKGP has a matrix product steady state, which can also be written as a linear superposition of Bernoulli shock measures~\cite{KJS}, and that it exhibits a phase transition at  $w_{1}=w_{2}$ from a low-density to a high-density phase. Using the two-dimensional matrix representation introduced in~\cite{Jafarpour}, one can calculate the average entropy production rate in the limit $L\to\infty$ and to leading order in $T$, obtaining
\begin{equation}
\langle \dot{S}_{\rm env} \rangle=
\left\{
\begin{array}{ll}
\frac{2\alpha w_{1}}{\alpha+w_{1}-w_{2}}\ln T   
& \mbox{for} \; w_{1} > w_{2}, \\ & \\
\frac{2 \beta w_{2}}{\beta+w_{2}-w_{1}} \ln T
& \mbox{for} \; w_{2} > w_{1}.
\end{array}
\right.
\end{equation}
Since the probability of a given configuration consisting of a particle in front of an empty lattice site is zero in the steady state, the last three processes in (\ref{RULESAKGP}) do not contribute to the average entropy production, explaining why the parameters $\omega_3,\omega_4,\omega_5$ do not appear in the result. In Fig.~\ref{fig:akgp} the average entropy production rate is plotted as a function of $w_{2}$. In contrast to the BCP, it changes \textit{continuously} at the transition point.

%==========================================================================
\section{Conclusions}
%==========================================================================

In this Letter we have addressed the problem of entropy production in systems with irreversible transitions. For such systems the Schankenberg formula~(\ref{EntropyProduction}) predicts an infinite entropy production which is not seen in experiments. We suggest that the finite amount of entropy produced in experiments is related to the fact that vanishing reverse rates are impossible in nature, it is only possible to keep such rates very small. By introducing the concept of a `defining rate' and an `actual rate' and estimating the latter by Bayesian inference, we could specify a lower bound on the entropy production which splits up into a constant contribution for reversible transitions and an additional contribution for irreversible transitions which grows logarithmically with the observation time $T$. 

Using the modified entropy production formula, we have calculated the average entropy production rate for three exactly solvable reaction-diffusion models in the steady state (further examples are discussed in the Supplemental Material~\cite{SM}). The steady state of the BCP and AKGP are very similar in the sense that they both can be written as a linear superposition of Bernoulli shock measures, where the shock performs a simple random walk~\cite{KJS}. However, the average entropy production behaves quite differently, namely, discontinuously in the BCP and continuously in the AKGP. This suggests that the irreversible entropy production may be used as an additional tool for the classification of nonequilibrium phase transitions.
%==========================================================================


\section*{References}
\begin{thebibliography}{99}
%==========================================================================

\bibitem{Schnakenberg}
Schnakenberg J, \textit{Network theory of microscopic and macroscopic behavior of master equation systems,} 1976 Rev. Mod. Phys. {\bf 48}, 571. 

\bibitem{Gaspard}
Andrieux D and Gaspard P, \textit{Fluctuation theorem and Onsager reciprocity relations}, 2004 J. Chem. Phys. {\bf 121} 6167.

\bibitem{Seifert}
Seifert U, \textit{Entropy production along a stochastic trajectory and an integral fluctuation theorem}, 2005 Phys. Rev. Lett {\bf 95}, 040602.

\bibitem{Seifert2}
Seifert U and Speck T, \textit{Fluctuation-dissipation theorem in nonequilibrium steady states}, Europhys. Lett. {\bf 89}, 10007 (2010).

\bibitem{Kolkata}
Hinrichsen H, Gogolin C, and Janotta P, \textit{Non-equilibrium dynamics, thermalization and entropy production}, 2011 J. Phys.: Conf. Ser. \textbf{297}, 012011.

\bibitem{SM}
see attached supplemental material.

\bibitem{Korda}
Korda PT,  Taylor MB, and  de Grier G, \textit{Kinetically locked-in colloidal transport in an array of optical tweezers}, Phys. Rev. Lett. \textbf{89}, 128301 (2002).

\bibitem{Lacasta}
Lacasta AM, Sancho JM, Romero AH, and Lindenberg K, \textit{Sorting on Periodic Surfaces}, Phys. Rev. Lett. \textbf{94}, 160601 (2005).

\bibitem{Dickman}
de Oliveira Rodrigues JE and Dickman R, \textit{Asymmetric exclusion process in a system of interacting Brownian particles}, Phys. Rev. E \textbf{81}, 061108 (2010). 

\bibitem{Haye}
see e.g. Hinrichsen H, \textit{Nonequilibrium phase transitions}, 2000 Adv. Phys. \textbf{49}, 815.

\bibitem{Takeuchi}
Takeuchi KA, Kuroda M, Chat{\'e} H, and Sano M, \textit{Directed percolation criticality in turbulent liquid crystals}, 2007 Phys. Rev. Lett. 99, 234503.

\bibitem{benAvraham}
ben-Avraham D, Dorosz S, and Pleimling M, \textit{Entropy production in nonequilibrium steady states: A different approach and an
exactly solvable canonical model}, 2011 Phys. Rev. E 84, 011115.

\bibitem{Bayes}
see e.g. Box GEP and Tiao GC, \textit{Bayesian Inference in Statistical Analysis}, Wiley and Sons, New York (1990).

\bibitem{DEHP}
Derrida B, Evans MR, Hakim V, and Pasquier V, \textit{Exact correlation functions in an asymmetric exclusion model with open boundaries}, 1993 J. Phys. A: Math. Gen. {\bf 26} 1493.

\bibitem{KJS} 
Krebs K, Jafarpour FH, and Sch\"utz GM, \textit{Microscopic structure of travelling wave solutions in a class of stochastic interacting particle systems}, 2003 New J. Phys. {\bf 5} 145.

\bibitem{Jafarpour}
Jafarpour FH and Masharian SR, \textit{Matrix product steady states as superposition of product shock measures in 1D driven systems}, 2007 J. Stat. Mech: Theor. Exp. P10013.




\end{thebibliography}
\end{document}

\title[Entropy Production of Nonequilibrium Steady States with Irreversible Transitions]{Supplemental Material for\\
``Entropy Production of Nonequilibrium Steady States with Irreversible Transitions''}
\author{Somayeh Zeraati$^1$, Farhad H. Jafarpour$^1$, and Haye Hinrichsen$^2$}
\address{$^1$Bu-Ali Sina University, Physics Department, 65174-4161 Hamedan, Iran}
\address{$^2$Universit\"at W\"urzburg, Fakult\"at f\"ur  Physik und Astronomie, 97074  W\"urzburg, Germany.}

\begin{abstract}
In this collection of Supplemental Material we recall the derivation of the formula for entropy production, provide further details of the calculations, and discuss the physical meaning of infinite entropy production for irreversible rates. Moreover, we present further explicit examples to demonstrate how the proposed concept works in systems with irreversible transitions.
\end{abstract}

\def\d{{\rm d}}
\def\0{\emptyset}

\parskip 2mm

\def\headline#1{{\noindent\bf #1: }}

\def\clr {\color{red}}
\def\clrb {\bf \color{red}}
\def\0{\emptyset}
\def\d{{\rm d}}
\def\hrev{ \dot s_{\rm env}^{\rm rev} }
\def\hirr{ \dot s_{\rm env}^{\rm irr} }

\tableofcontents
\newpage
\parskip 2mm
\title[Entropy Production of Nonequilibrium Steady States with Irreversible Transitions]{}
%================================================================================
\section{Entropy production in nonequilibrium steady states}
%================================================================================

In classical statistical mechanics, complex systems are often modeled as continuous-time Markov processes. Such systems evolve by spontaneous random transitions from configuration $c$ to configuration $c'$ according to certain rates $w_{c\to c'}$. While the specific trajectory of such a stochastic process is unpredictable, the probability $P_c(t)$ to find the system at time $t$ in configuration $c$ evolves deterministically according to the master equation~\cite{vanKampen}
\begin{equation}
\frac{\partial}{\partial t} P_c(t) 
\;=\; \sum_{c'\in\Omega} \Bigl(J_{c'\to c}(t)-J_{c \to c'}(t)\Bigr)\,,
\end{equation}
where $J_{c \to c'}(t) \;=\; P_c(t)\, w_{c \to c'}$ denotes the probability current flowing from configuration $c$ to configuration $c'$. 

\begin{figure}[t]
\centering\includegraphics[width=60mm]{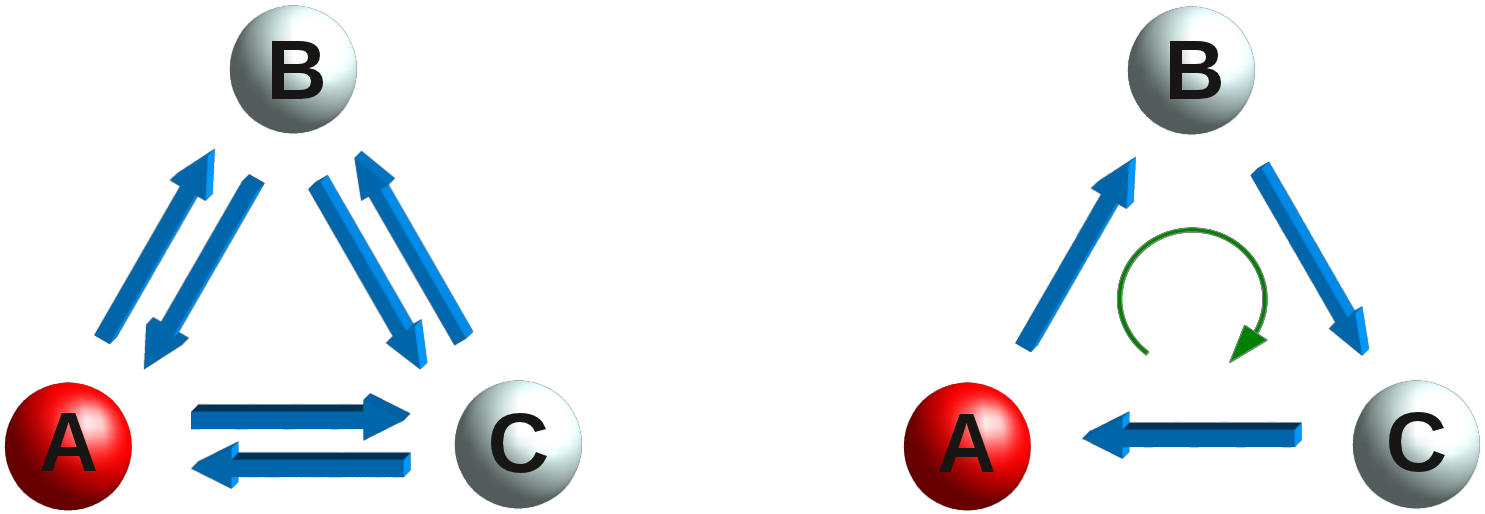}
\caption{Markov process with three configurations $A,B,C$. The system is currently in state $A$ and hops randomly as indicated by the arrows. Left: For symmetric rates the system performs an unbiased random walk in its state space and evolves into a stationary equilibrium state with $P_A=P_B=P_C=1/3$ which obeys detailed balance. Right: For totally asymmetric rates the system evolves into a stationary state with the same probabilities. However, as the systems can only move clockwise, this state is characterized by nonvanishing probability currents and therefore it is out of equilibrium. In nature such a NESS requires that the system is constantly driven from outside. \label{fig:clock}}
\end{figure}

A system is called stationary if its probability distribution $P_c(t)$ and therewith the currents $J_{c \to c'}(t)$ are time-independent. Moreover, if a stationary system obeys the additional condition of detailed balance
\begin{equation}
J_{c \to c'} = J_{c' \to c}\,, \qquad \forall c,c'\,,
\end{equation}
meaning that all microscopic probability currents cancel pairwise, it is in a thermal equilibrium state. Otherwise, if the condition of detailed balance is violated, the system is said to be in a nonequilibrium steady state (NESS)~\cite{NESS}. In this case the system performs a biased random walk in its own configuration space, leading to nonvanishing probability currents  as illustrated in Fig.~\ref{fig:clock}.

In nature nonequilibrium steady states cannot exist on their own, instead they can only exist in systems which are continuously driven from outside. In experiments this external drive usually manifests itself as a physical current such as a flow of heat or particles. On the level of the model the external drive is not implemented explicitly, it is rather encoded in the asymmetry of the rates $w_{c\to c'}$. 

While the internal entropy of a system in a NESS is on average constant, the external drive will continually produce entropy in the environment~\cite{Gallavotti}. Remarkably, there exists a theoretical result which allows one to quantify this external entropy production without knowing the specific physical properties of the environment. In fact, based on previous works by Schnakenberg~\cite{Schnakenberg}, Andrieux and Gaspard~\cite{Gaspard}, it was shown by Seifert~\cite{Seifert,Seifert2} that the environmental entropy production along a stochastic path of transitions $c_0 \to c_1 \to \ldots$ at times $t_0,t_1,\ldots$ is given by
\begin{equation}
\label{SeifertsEntropy}
\dot{S}_{\rm env}  \;=\;\sum_{j} \delta(t-t_{j}) \ln\frac{w_{c_j\to c_{j+1}}}{w_{c_{j+1}\to c_j}}\,.
\end{equation}
%
%======================================================
%======================================================
%
In a numerical simulation of a given model, where the rates are known, the entropy production can be measured very easily by adding up all discontinuous changes
\begin{equation}
 \label{EntropyChange}
 \Delta S_{\rm env}(c \to c') \;=\; \ln \frac{w_{c\to c'}}{w_{c'\to c}}\,
\end{equation}
whenever the system hops from $c$ to $c'$. In the long-time limit these entropy fluctuations obey the fluctuation theorem $P(-\Delta S)=e^{-\Delta S}P(\Delta S)$ which implies that the entropy production is on average positive. Averaging Eq.~(\ref{SeifertsEntropy}) over many possible paths one obtains the mean entropy production rate
 \begin{equation}
 \label{AverageEntropyProduction}
\langle \dot{S}_{\rm env} \rangle = \sum_{c \neq c'} P_cw_{c \to c'}\Delta S_{\rm env}(c \to c')\,,
 \end{equation}
which can be evaluated if the stationary probabilities $P_c$ are known~\cite{NESSEntropy,Tome}. In lattice models the entropy production usually scales with the number of sites $L$, allowing us to define the average entropy production rate per site
\begin{equation}
\dot s_{\rm env} = \frac1L \langle \dot S_{\rm env} \rangle\,.
\end{equation}

%================================================================================
\section{Motivating the entropy production formula}
%================================================================================

\noindent
For readers who are not familiar with the concept of entropy production we start with a derivation in the limit of fast equilibration in the environment, following a more extended presentation given in Ref.~\cite{Kolkata}.

\headline{Isolated systems}
%---------------------------------
In classical statistical physics a complex system is usually modeled as a set $\Omega$ of configurations $c\in\Omega$, in which the system is assumed to hop spontaneously between different configurations according to certain rates $w_{c \to c'}$. Starting point of equilibrium statistical physics is the \textit{equal a priori postulate}, which states that an \textit{isolated} system always evolves into a stationary state of maximal entropy, meaning that all configurations are visited with the same probability $P_c=1/|\Omega|$. Moreover, as the underlying closed quantum system is invariant under time reversal, one can show that the stationary state of an isolated system obeys detailed balance, meaning that the rates of an isolated system are always symmetric, i.e. $ w_{c \to c'}=w_{c' \to c}$. 

\headline{Open systems}
%---------------------------------
If a system interacts with its environment, the system together with its environment can be considered in itself as a large isolated system (`Universe') which relaxes into a state of maximal entropy. Let us denote the configurations of this combined system by $u \in \Omega_{\rm tot}$ and the configurations of the contained laboratory system as $c \in \Omega_{\rm sys}$. Clearly, for a given state $c$ of the laboratory system the environment can be in many different possible states.  This means that there exists a projection $\pi:\Omega_{\rm tot}\mapsto \Omega_{\rm sys}$, mapping each state $u$ of the total system onto the corresponding state $c=\pi(u)$ of the laboratory system. 

\begin{figure}[t]
\begin{flushright}
\includegraphics[width=130mm]{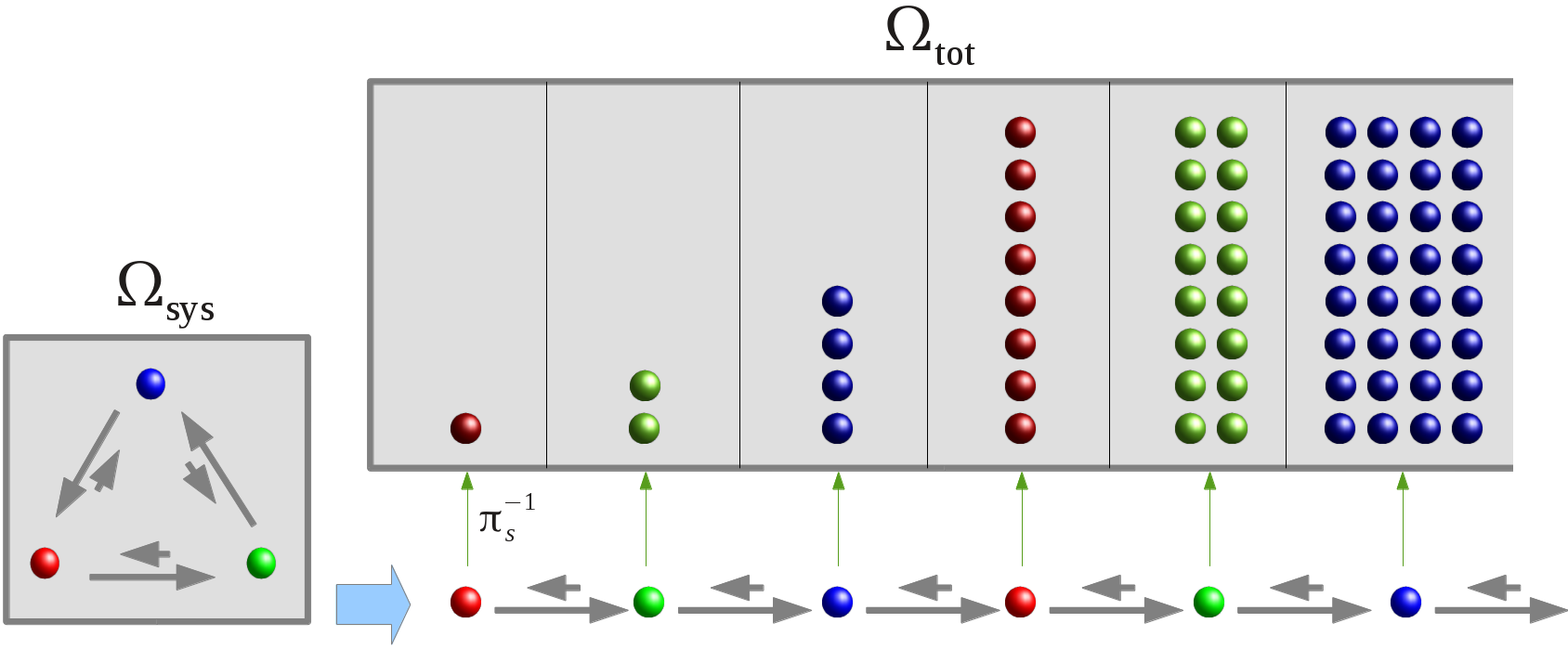}
 \end{flushright}
\caption{Illustration of a nonequilibrium system with three configurations (red, green and blue) and biased rates. A given configuration, say the red one, is compatible with many different red states in the total system which are arranged in dynamically separated sectors. To move from one sector to the other, the laboratory system has to change its configuration. The effective bias of the rates in the laboratory system is caused by an ever increasing number of the corresponding available configurations in the respective sector of the total system.\label{fig:sectors}}
\end{figure}

\headline{Dynamically connected sectors}
%---------------------------------
If the laboratory system is in configuration $c$, the configuration of the total system will belong to the subspace $\pi^{-1}(c)=\{u\in \Omega_{\rm sys}|\pi(u)=c\}$. As illustrated in Fig.~\ref{fig:sectors}, this subspace generally decomposes into many dynamically separated sectors. As long as the laboratory system stays in $c$, the environment can only evolve within the actual sector, while a change to a different sector requires the laboratory system to evolve through a closed cycle from $c$ over various intermediate configurations back to $c$. In the following we denote by $\pi^{-1}_s$ the current sector of dynamically connected system states which are compatible with the current configuration $c$ for a given temporal evolution in the past.

%======================================================
%======================================================

\headline{Reduced rates}
%---------------------------------
%
The probability distribution $P_u(t)$ of the full system consisting of laboratory system and environment evolves according to the master equation
\begin{equation}
\label{FullMaster}
\frac{\d}{\d t} P_u(t) \;=\; \sum_{u'\in\Omega_{\rm tot}} \Bigl( J_{u' \to u}(t)-J_{u \to u'}(t)\Bigr)\,,
\end{equation}
where $J_{u \to u'}(t)=P_u(t)w_{u \to u'}$ denotes the probability current flowing from $u$ to $u'$. Since the total system is assumed to be isolated, the rates have to be time-independent and symmetric. i.e.
\begin{equation}
w_{u \to u'}=w_{u' \to u}\,.
\end{equation}
Likewise, the probability distribution $P_c(t)$ of the laboratory system evolves according to a master equation
\begin{equation}
\label{ReducedMaster}
\frac{\d}{\d t} P_{c}(t) \;=\; \sum_{c'\in\Omega_{\rm sys}} \Bigl( J_{c' \to c}(t)-J_{c \to c'}(t)\Bigr)
\end{equation}
with the probability currents $J_{c \to c'} =P_c(t)w_{c \to c'}(t)$. Obviously, this reduced evolution equation is only compatible with the equation (\ref{FullMaster}) of the total system if 
\begin{equation}
P_c(t) =  \hspace{-3mm} \sum_{u \in \pi_s^{-1}(c)} \hspace{-2mm} P_u(t)\,,\qquad
J_{c \to c'}(t) =\hspace{-4mm} \sum_{{u \in \pi_s^{-1}(c)} \atop {{u' \in \pi_s^{-1}(c')}}} \hspace{-3mm}J_{u \to u'}(t).
\end{equation}
This implies that the reduced rates $w_{c \to c'}(t)$ in the laboratory system are given by
\begin{equation}
\label{reducedrates}
w_{c\to c'}(t) = \frac{J_{c \to c'}(t)}{P_c(t)}
= \frac{ \sum_{{u \in \pi_s^{-1}(c)} \atop {{u' \in \pi_s^{-1}(c')}}} J_{u \to u'}(t)}
{\sum_{u \in \pi_s^{-1}(c)} P_u(t)} \,.
\end{equation}
It is important to note that the reduced rates are generally time-dependent. However, experiments are often designed in such a way that the rates are effectively kept constant during the observation time.

\headline{Separation of time scales}
%-----------------------------------------
%
The easiest way to derive the formula for entropy production $ \Delta S_{\rm env} = \ln \frac{w_{c\to c'}}{w_{c'\to c}}$ is to assume that the environment is in thermal equilibrium and equilibrates almost instantaneously within its dynamically accessible sector whenever the laboratory system changes its configuration. This means that for a given configuration $c$ of the laboratory system, all configurations $u \in \pi_s^{-1}(c)$ are equally probable, i.e.
\begin{equation}
\label{equalprobs}
P_u(t)=\frac{P_c(t)}{N_c(t)} \qquad \forall u\in \pi_s^{-1}(c)
\end{equation}
where $N_c(t) = |\pi_s^{-1}(c)|$ is the number of states in the current dynamically accessible sector in the total system. Inserting this equation into Eq.~(\ref{reducedrates}), the reduced rates are given by
\begin{equation}
w_{c\to c'}(t) = \frac{1}{N_c(t)}\sum_{{u \in \pi_s^{-1}(c)} \atop {{u' \in \pi_s^{-1}(c')}}} w_{u \to u'}
\end{equation}
so that the quotient of mutually reverse transition rates can be expressed by a ratio of the number of states
\begin{equation}
\frac{w_{c\to c'}(t)}{w_{c'\to c}(t)} \;=\; \frac{N_{c'}(t)}{N_c(t)}\,,
\end{equation}
where we used the symmetry $w_{u \to u'}=w_{u' \to u}$ of the rates in the (isolated) total system.\\

\headline{Entropy production in the environment}
%-----------------------------------------
%
In the limit of a quasi-instantaneous equilibration the entropy in the environment for a given $c$ is simply given by
\begin{equation}
S_{\rm env}(c) \;=\; \ln N_c(t)\,.
\end{equation}
Whenever the system jumps from $c \to c'$, the environmental entropy thus changes by
\begin{equation}
\label{DeltaSEnv}
\Delta S_{\rm env}(c \to c') \;=\; \ln \frac {N_{c'}(t)}{N_c(t)} \;=\; \ln\frac{w_{c \to c'}(t)}{w_{c' \to c}(t)}\,.
\end{equation}
Along a stochastic path $\gamma: c_0 \to c_1 \to \ldots$ of transitions taking place at $t_0,t_1,\ldots$, the temporal derivative of the entropy in the environment is therefore given by~\cite{Seifert}
\begin{equation}
\label{entropyproduction}
\dot S_{\rm env} \;=\;  \sum_j \delta(t-t_j) \ln \frac{w_{c_j \to c_{j+1}}(t)}{w_{c_{j+1} \to c_j}(t)}
\end{equation}
which reproduces formula (3) in the Letter. Averaging~(\ref{entropyproduction}) over all possible paths gives the average entropy production rate in Eq. (\ref{AverageEntropyProduction}):
\begin{equation}
\langle \dot{S}_{\rm env} \rangle = \sum_{c \neq c'} P_c(t)w_{c \to c'}(t)\,\Delta S_{\rm env}(c \to c').
\end{equation}
If the laboratory system is stationary, the $t$-dependence on the r.h.s. can be dropped.\\

\headline{The problem of infinite entropy production}
%-----------------------------------------
%
For irreversible transitions, where $w_{c_j \to c_{j+1}}>0$ and $w_{c_{j+1} \to c_j}=0$, the logarithm in Eq.~(\ref{entropyproduction}) diverges, meaning that such a transition would instantaneously produces an infinite amount of entropy. In fact, in the setting discussed above, an irreversible transition would correspond to an infinite increase of the number of states in the environment. However, in the laboratory we can realize irreversible processes experimentally without seeing an infinite entropy production, which would result into a divergent increase of heat in the environment. In the Letter we argue that this apparent paradox can be overcome by realizing that microscopically irreversible rates are in principle impossible in nature. However, we can design an experiment in such a way that the reverse transition so strongly suppressed that it practically never occurs during the life time $T$ of the experiment. However, observing that a particular transition does not occur in an experimental realization 
during a finite time $T$ of data taking, we cannot conclude that the actual rate was exactly zero, all what we can say is that this rate of the order of $1/T$ or smaller. Therefore we suggest that the actual rate in the experiment is small but positive, leading to a finite entropy production with a lower bound that grows logarithmically with~$T$. The slow logarithmic dependence in $T$ explains why we practically do not see an infinite entropy production in nature.

The argument can also be spelled out in a different way: In nature it is in principle impossible to design experiments with irreversible processes. However, we can approximate irreversible processes quite well in experiments, but the effort to do this (reflected in the entropy production) grows only logarithmically with the timescale of the rate which is sent to zero. In other words, the better we approximate a zero rate experimentally, the more entropy the experimental setup will generate in the environment, but the entropy grows so slowly that a divergence in the limit of a vanishing rate is not seen in practice.

%================================================================================
\section{Full derivation of the entropy production for irreversible processes}
%================================================================================
%
In the paper we presented a simplified derivation of the irreversible entropy production, where we used the mathematically incorrect replacement $\langle \ln w \rangle \to \ln \langle w \rangle$. Here we present the full derivation, showing that in the present case one obtains essentially the same result.

As in the main text we consider a particular pair of states $c,c'$ with a defining rate $w=w_{c\to c'}$ and denote by $\tau=P_{c}T$ the average time spent in configuration $c$ during data taking. With the likelihood $P(n| \tilde w)=(\tau \tilde w)^ne^{-\tau \tilde w}/n!$ and its conjugate prior $P(\tilde w)={\tilde\beta}^{\tilde\alpha} \tilde w^{{\tilde\alpha}-1} e^{-{\tilde\beta} \tilde w}/\Gamma({\tilde\alpha})$ the normalizing marginal likelihood $P(n)$ in Bayes' rule
\begin{equation}
\label{Bayes}
P(\tilde w|n)=\frac{P(n|\tilde w)P(\tilde w)}{P(n)} 
\end{equation}
is given by
\begin{eqnarray}
P(n)&=&\int_0^\infty\d \tilde w\,P(n|\tilde w)P(\tilde w) \\ \nonumber &=& \frac{\tau^n {\tilde\beta}^\alpha\, \Gamma(n+\tilde\alpha)}{ (\tau+\tilde\beta)^{n+\tilde\alpha}\, \Gamma(1+n) \Gamma(\tilde\alpha)}\,.
\end{eqnarray}
Inserting this result into Eq.~(\ref{Bayes}) we obtain the posterior
\begin{equation}
P(\tilde w|n) \;=\; \frac{\tilde w^{\tilde\alpha +n-1} e^{-(\tilde\beta+\tau)\tilde w} (\tilde\beta +\tau)^{\tilde\alpha +n}}{\Gamma
   (n+\tilde\alpha )}\,,
\end{equation}
which describes the expected distribution of the actual rate $\tilde w$ for a given number of events $n$ during the observation time $T$ under the assumption that the rates are distributed according to the prior with certain hyperparameters $\tilde\alpha,\tilde\beta$. In the Letter we first compute the expectation value of the actual rate
\begin{equation}
\label{Average}
\langle \tilde w \rangle = \int \d \tilde w \, \tilde w \sum_{n=0}^\infty P(\tilde w|n)P(n|w)= \frac{\tau w+\tilde\alpha}{\tau+\tilde\beta}\,
\end{equation}
for which one can easily show that the integration and the infinite sum commute. This result is then plugged into the entropy production formula for the actual rates
\begin{equation}
\Delta S_{\rm env} (c \to c')= \ln \frac{ \langle \tilde{w}_{c\to c'}\rangle}{  \langle \tilde{w}_{c'\to c}\rangle}\,,
\end{equation}
giving
\begin{equation}
\label{EP1}
\Delta S_{\rm env} =
 \left\{\begin{array}{cc}  
\ln \frac{ w_{c\to c'} }{ w_{c'\to c} } & \mbox{for reversible transitions,} \\[4mm]
\ln \frac{ T P_{c'}  w_{c\to c'} }{ \tilde\alpha}  & \mbox{for irreversible transitions.} \end{array} \right.
\end{equation}
for large observation time $T \gg 1$. However, this simplified calculation is incorrect since we have replaced
\begin{equation}
\langle \ln w \rangle \to \ln \langle w \rangle
\end{equation}
in the entropy production formula, but the arithmetic average does not commute with a nonlinear function. 
In a correct derivation, the logarithm should be taken \textit{before} averaging over $\tilde w$ and $n$. 
In fact, using {\textit{Mathematica}\texttrademark} one can show that
\begin{equation*}
\langle \ln \tilde w | n \rangle = \int_0^\infty \d\tilde w \, \ln \tilde w\, P(\tilde w | n) = \psi(n+\tilde\alpha)-\ln(\tau+\tilde\beta)\,,
\end{equation*}
where $\psi(z)$ denotes the digamma function. This leads to
\begin{eqnarray}
\langle \ln \tilde w \rangle &=& \sum_{n=0}^\infty \langle \ln \tilde w | n \rangle P(n|w)  \\ \nonumber
&=& \psi(\tilde\alpha)-\ln(\tau+\tilde\beta)-\left.\frac{\partial}{\partial a} {_1F_1}(a;\tilde\alpha;-\tau w)\right|_{a=0}\,,
\end{eqnarray}
where ${_1F_1}(a;b;z)$ is the Kummer confluent hypergeometric function. Again one can show that the integral over $\tilde w$ and the infinite sum over $n$ commute. 
By defining $\tau'=P_{c'}T$ and inserting this result into the entropy production formula we obtain
\begin{eqnarray}
\Delta S_{\rm env} &=& \langle \ln \tilde w_{c \to c'} \rangle - \langle \ln \tilde w_{c' \to c} \rangle \\ \nonumber 
&=& -\Big( \ln(\tau+\tilde\beta)-\ln(\tau'+\tilde\beta) \Big) \\ \nonumber
&&+\frac{\partial}{\partial a} \Big(-{_1F_1}(a;\tilde\alpha;-\tau w_{c \to c'}) \left.+{_1F_1}(a;\tilde\alpha;-\tau' w_{c' \to c})\Bigr)\,\right|_{a=0}\,.
\end{eqnarray}
For $a>0$ the Kummer confluent hypergeometric function has the integral representation
\begin{equation}
{_1F_1}(a;b;z)=\frac{\Gamma(b)}{\Gamma(a)\Gamma(b-a)}\int_0^1 e^{zu}u^{a-1}(1-u)^{b-a-1}\,\d u
\end{equation}
with the derivative
\begin{eqnarray}
&&\frac{\partial}{\partial a}{_1F_1}(a;b;z)=\frac{\Gamma(b)}{\Gamma(a)\Gamma(b-a)}\int_0^1 e^{zu}u^{a-1}(1-u)^{b-a-1}\nonumber\\  
&& \qquad  \times \Bigl(\psi(b-a)-\psi(a)-2 \tanh ^{-1}(1-2 u)\Bigr)\,\d u\,.
\end{eqnarray}
While each of the Kummer functions in this representation would diverge when taking $a \to 0$, it turns out that their difference remains finite. The resulting integral is given by
\begin{equation}
\label{EntropyIntegral}
 \Delta S_{\rm env} \;=\; \ln \frac{\tau'+\tilde\beta}{\tau+\tilde\beta}+\int_0^1 \frac{(e^{-u \tau' w_{c'\to c}}-e^{-u \tau w_{c \to c'}})}{u(1-u)^{1-\tilde\alpha}}\d u.
\end{equation}
This integral can be evaluated for an arbitrary $\tilde\alpha$. For large observation time $T\gg 1$ the entropy production given by~(\ref{EntropyIntegral}) behaves asymptotically as
\begin{equation}
\label{EP2}
\Delta S_{\rm env} =
 \left\{\begin{array}{cc}  
\ln \frac{w_{c\to c'}}{w_{c'\to c}} & \mbox{for reversible transitions,} \\[4mm]
\ln \frac{T P_{c'}  w_{c\to c'}}{\tilde\alpha_R}  & \mbox{for irreversible transitions.} \end{array} \right.
\end{equation}
where $\tilde \alpha_R = \exp (\psi(\tilde\alpha))$ replaces the $\tilde\alpha$-hyperparameter.  Hence, by commuting the logarithm and the average, we obtain to leading order the same result with a redefined hyperparameter $\tilde\alpha$.

For the special case $\tilde\alpha=1$ the entropy production~(\ref{EntropyIntegral}) can be calculated exactly and results in
\begin{equation}
\Delta S_{\rm env} \;=\; \ln \frac{\tau'+\tilde\beta}{\tau+\tilde\beta}+\Gamma(0,\tau w_{c \to c'}) - \Gamma(0,\tau' w_{c' \to c}) + \ln \frac{P_{c}w_{c \to c'}}{P_{c'}w_{c' \to c}}
\end{equation}
in which $\Gamma(s,z)$ is the incomplete gamma function. For reversible transitions, the incomplete gamma function $\Gamma(0,\tau w)$ vanishes exponentially in the limit $T \to \infty$ and can be neglected, hence the calculation reproduces the usual entropy production formula. For irreversible transitions with $w_{c' \to c}=0$, however, we obtain
\begin{equation}
\Delta S_{\rm env} \;=\; \ln \frac{\tau'+\tilde\beta}{\tau+\tilde\beta}+\gamma + \Gamma(0,\tau w_{c \to c'}) +  \ln(\tau w_{c \to c'})\,,
\end{equation}
where $\gamma$ is Euler's constant. In the long time limit the incomplete gamma function goes to zero exponentially, hence the first and the last terms give the entropy production for irreversible transitions.

%================================================================================
\section{Entropy production in systems with matrix product steady states}
%================================================================================
%
There are families of one-dimensional stochastic nonequilibrium systems, such as driven diffusive systems, for which the exact steady state probability distribution can be expressed in a matrix product form. In what follows we briefly review the matrix product method (for a complete review see~\cite{BE}) and show how we can calculate the entropy production in a simple one-dimensional stochastic nonequilibrium system. 

Let us consider a one-dimensional lattice with $L$ lattice sites and open boundaries. To each lattice site $i$ we assign an occupation number $c_{i}$ associated with different states of that lattice site. The idea is to associate an operator with each possible state of a lattice site and write the unnormalized steady state weight of a given configuration $c =\{ c_{1},\cdots, c_{L} \}$ for periodic boundary conditions as
\begin{equation}
\label{MPSP}
P_c \propto Tr(D_{c_{1}}D_{c_{2}}\cdots D_{c_{L}}).
\end{equation} 
and for open boundaries as
\begin{equation}
\label{MPS}
P_c \propto \langle W \vert D_{c_{1}}D_{c_{2}}\cdots D_{c_{L}} \vert V \rangle.
\end{equation}
with appropriate vectors $\vert V \rangle$ and $\langle W\vert$. Requiring~(\ref{MPS}) to satisfy the master equation~(\ref{ReducedMaster}) in the steady state 
\begin{equation}
\sum_{ c' \neq c } P_{c} w_{c \to c'} = \sum_{c' \neq c} P_{c'} w_{c' \to c}
\end{equation}
determines the relation between the noncommuting operators and the vectors $\vert V \rangle$ and $\langle W \vert$. 

As a simple example let us consider the Totally Asymmetric Simple Exclusion Process (TASEP) with open boundaries~\cite{DEHP}. The dynamical rules are
\begin{equation}
\label{TASEP}
\begin{array}{l}
1 \0 \to \0 1\;\mbox{with rate}\;1\\
\0 \to 1       \;\mbox{at the left boundary with rate}\;\alpha\\
1  \to \0      \;\mbox{at the right boundary with rate}\;\beta.
\end{array}
\end{equation}
As a two-state system we associate the operators $D_0=E$ and $D_1=D$ with an empty lattice site ($c_{i}=0$) and an occupied lattice site ($c_{i}=1$) respectively. These operators besides the vectors $\vert V \rangle$ and $\langle W \vert$ satisfy the following quadratic algebra~\cite{DEHP}
\begin{equation}
\label{ASEPALG}
\begin{array}{l}
DE=D+E\\ \\
D\vert V \rangle= \frac{1}{\beta}\vert V \rangle \\ \\
\langle W \vert E=\frac{1}{\alpha} \langle W \vert.
\end{array}
\end{equation}
It is known that the algebra~(\ref{ASEPALG}) has an infinite-dimensional matrix representation. Note that all of the transitions in the TASEP are irreversible, hence using~(\ref{AverageEntropyProduction}) and~(\ref{EP2}) the average entropy production to leading order in $T$ can be written as follows
\begin{eqnarray}
\label{TASEPEntropy}
\langle \dot S_{\rm env} \rangle & \approx & \Big ( \sum_{\{c\}}\alpha P_{\{0 c_2 \cdots c_{L}\}} + 
\sum_{i=1}^{L-1}\sum_{\{c\}}P_{\{c_1 c_2 \cdots c_{i-1} 10 c_{i+2} \cdots c_{L}\}} \nonumber \\
& + & \sum_{\{c\}}\beta P_{\{c_1 \cdots c_{L-1}1\}} \Big ) \ln T.
\end{eqnarray}
The first and the last terms in~(\ref{TASEPEntropy}) are contributions of the boundaries in the average entropy production while the second term comes from the diffusion of particles in the bulk of the lattice. Using the algebra~(\ref{ASEPALG}) the corresponding probabilities in~(\ref{TASEPEntropy}) can be calculated 
\begin{eqnarray}
&&\sum_{\{c\}}P_{\{0 c_2 \cdots c_{L}\}}=\frac{\langle W \vert EC^{L-1}\vert V \rangle}{Z_{L}}=\frac{1}{\alpha}\frac{Z_{L-1}}{Z_{L}}\\
&&\sum_{i=1}^{L-1}\sum_{\{c\}}P_{\{c_1 c_2 \cdots c_{i-1} 10 c_{i+2} \cdots c_{L}\}}\\ \nonumber
&&=
\sum_{i=1}^{L-1}\frac{\langle W \vert C^{i-1} D E C^{L-i-1}\vert V \rangle}{Z_{L}}=(L-1)\frac{Z_{L-1}}{Z_{L}}\\
&&\sum_{\{c\}}P_{\{c_1 \cdots c_{L-1}1\}}=\frac{\langle W \vert C^{L-1} D\vert V \rangle}{Z_{L}}=\frac{1}{\beta}\frac{Z_{L-1}}{Z_{L}}.
\end{eqnarray}
in which we have defined $C=E+D$. The normalization factor $Z_{L}$ is given by~\cite{DEHP}
\begin{equation}
Z_{L}=\langle W \vert C^L \vert V \rangle=\sum_{k=1}^{L}\frac{k(2L-k-1)!}{L!(L-k)!}\frac{\beta^{-k-1}-\alpha^{-k-1}}{\beta^{-1}-\alpha^{-1}}.
\end{equation}
The current of particles between the lattice sites $i$ and $i+1$ can be written as
\begin{equation}
J_{i,i+1}(\alpha,\beta)=\langle c_{i}(1-c_{i+1}) \rangle=\frac{\langle W \vert C^{i-1}DEC^{L-i-1}\vert V \rangle}{Z_{L}}.
\end{equation}
Using the algebra~(\ref{ASEPALG}) we can easily see that the stationary state current of particles is site-independent and is given by
\begin{equation}
J(\alpha,\beta)=\frac{Z_{L-1}}{Z_{L}}.
\end{equation}
In the thermodynamic limit $L\to\infty$ it is clear that the contribution of the bulk reaction in the average entropy production is dominant, resulting in
\begin{equation}
\frac{1}{L}\langle \dot S_{\rm env} \rangle \;\approx\; J(\alpha,\beta) \ln T
\end{equation}
which recovers the result obtained in the Letter.

Using the same procedure explained above one can, in principle, calculate the average entropy production of any stochastic nonequlibrium system given that its stationary state probability distribution can be obtained using the matrix product method. As it is mentioned in the Letter, we have used a two-dimensional matrix representations for the associated quadratic algebras of the Branching-Coalescing Process and the Asymmetric Kawasaki-Glauber Process to calculate the average entropy production in these systems. 

%================================================================================
\section{Further examples and applications}
%================================================================================

%================================================================================
\subsection{Chemical rate equations}
%================================================================================
Rate equations for chemical reactions describe the temporal evolution of concentrations of particles. For example, the Lotka-Volterra equations
\begin{eqnarray}
\dot {[X]}&=&\alpha[X]-\beta[X][Y]\\
\dot {[Y]}&=&-\gamma[Y]+\delta[X][Y]
\end{eqnarray}
describes the interaction of a prey population $[X]$ and a predator population $[Y]$ and exhibits oscillatory solutions for suitable parameters. 

Chemical evolution equations of this kind are mean field approximations in which the spatial dependence is ignored. They are often formulated as if the underlying chemical reactions were irreversible. For example, the Lotka-Volterra would correspond to the reaction scheme
\begin{equation}
X \rightarrow \alpha XX, \quad
XY \rightarrow \beta Y, \quad
Y \rightarrow \gamma \emptyset, \quad
XY \rightarrow \delta XYY.
\end{equation}
If no information about reverse rates are available, our result allows us to quantify the irreversible entropy production rate by
\begin{eqnarray}
\langle  \dot S_{\rm env} \rangle &=& 
[X]\alpha \ln (\frac{[X]^2\alpha T}{\tilde{\alpha}_R}) + 
[X][Y] \beta \ln (\frac{[Y]\beta T}{\tilde{\alpha}_R}) \\ \nonumber
&+& [Y] \gamma \ln(\frac{\gamma T}{\tilde{\alpha}_R}) + 
[X][Y] \delta \ln (\frac{[X][Y]^2\delta T}{\tilde{\alpha}_R}).
\end{eqnarray}
For an oscillatory solution the entropy production would oscillate as well.

%================================================================================
\subsection{Contact process}
%================================================================================

\begin{figure}[t]
\centering\includegraphics[width=105mm]{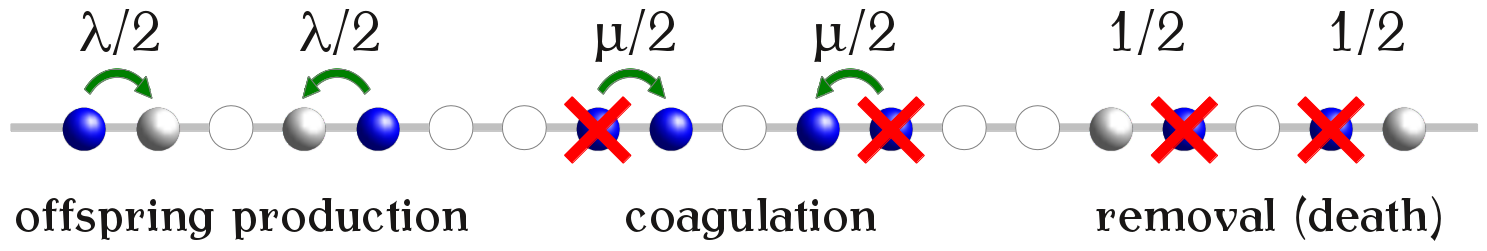}
\caption{Dynamical rules of the contact process. Offspring production and coagulation are reversible while the death process is irreversible. The model exhibits a nonequilibrium phase transitions.\label{fig:cp}}
\end{figure}

\begin{figure}[t]
\centering\includegraphics[width=120mm]{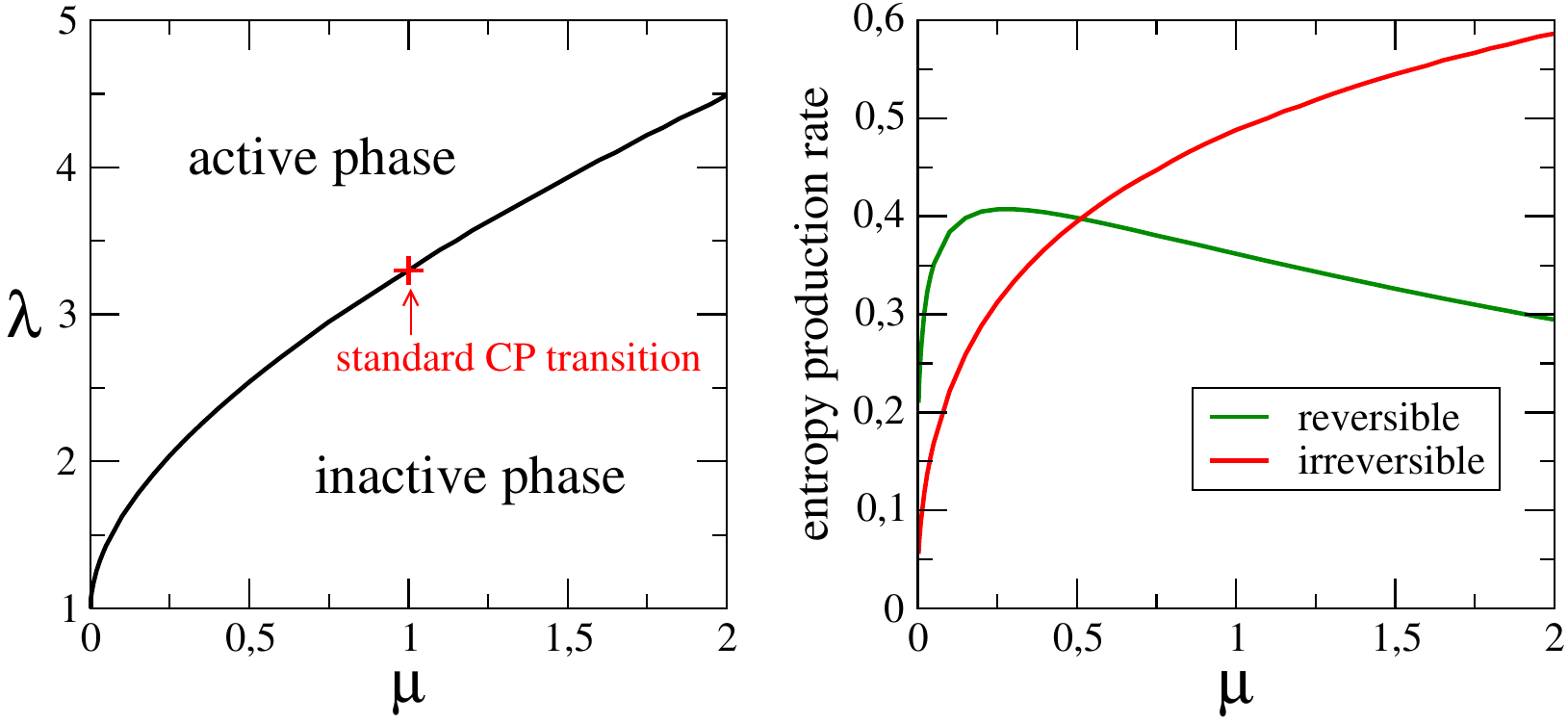}
\vspace*{-4mm}
\caption{Two-parameter contact process in 1+1 dimensions. Left: Phase diagram with a phase transition line. Right: Reversible and irreversible entropy production rate over $T=10$ Monte Carlo sweeps divided by the particle density along the transition line (see text). \label{fig:cpent}}
\end{figure}

The 1+1-dimensional contact process is a nonequilibrium process which is frequently used as a simple model for epidemic spreading~\cite{Harris,Liggett}. It evolves by random-sequential updates, i.e. a pair of adjacent sites is randomly selected and the following processes are carried out (cf. Fig.~\ref{fig:cp}):
\begin{eqnarray*}
\0 1\to 11  \qquad& \mbox{ with rate } \qquad & \lambda/2\\[-2mm]
1 \0\to 11  & \mbox{ with rate } & \lambda/2\\ 
11\to \0 1  & \mbox{ with rate } & \mu/2\\ [-2mm]
11\to 1\0   & \mbox{ with rate } & \mu/2\\ 
\0 1\to \0\0  & \mbox{ with rate } & 1/2\\ [-2mm]
1 \0\to \0\0  & \mbox{ with rate } & 1/2
\end{eqnarray*}
The standard version of the contact process in 1+1 dimensions with $\mu=1$ is known to exhibit an absorbing phase transition at the critical point $\lambda_c=3.29785(2)$~\cite{CPcrit}. In the two-parameter version defined above there is a critical line of transitions (see Fig.~\ref{fig:cpent}). Except for the lower terminal point at $\lambda=1, \mu=0$, where the dynamics is the same as in a Glauber-Ising model at zero temperature, all transition points along this line are expected to belong to the universality class of directed percolation (DP)~\cite{Haye}.

%======================================================
%======================================================

Although models in the DP class are nonintegrable, the entropy production of this model can be related exactly to the actual particle density. To see this let $P_{c_i,c_{i+1}}$ be the probability to find a randomly selected pair of sites $i,i+1$ in the states $c_i,c_{i+1}$. Clearly, these probabilities are left-right symmetric
\begin{equation}
\label{eqcp1}
P_{01}=P_{10}
\end{equation}
and normalized by
\begin{equation}
 P_{00}+P_{01}+P_{10}+P_{11}=1\,.
\end{equation}
Moreover, the average particle density can be computed by integrating out one of the sites, i.e.
\begin{equation}
 \rho = P_{00}+P_{10}\,.
\end{equation}
According to the dynamical rules defined above, the first four processes increase the particle number by one, whereas the last two processes decrease the particle number by one. Therefore, the density of particles will change in time as
\begin{equation}
\label{eqcp4}
\dot\rho = (P_{01}+P_{10})\frac{\lambda-1}2 - \mu P_{11}
\end{equation}
In the stationary state, where $\dot\rho=0$, Eqs. (\ref{eqcp1})-(\ref{eqcp4}) provide four equations, giving the two-site probabilities
\begin{equation}
P_{01}=P_{10}=\frac{\mu}{\lambda+\mu-1}\rho\,,\qquad 
P_{11} = \frac{\lambda-1}{\lambda+\mu-1}\rho
\end{equation}
which allow us to derive the following exact relations for the reversible and irreversible entropy production rate:
\begin{eqnarray*}
\hrev &=& \Bigl( \frac{P_{10}+P_{01}}{2}\lambda-P_{11}\mu \Bigr)\ln(\frac{\lambda}{\mu}) \;=\; \frac{\mu\,\rho}{\lambda+\mu-1}\ln(\lambda/\mu) ,\\
\hirr &=& \frac{P_{10}+P_{01}}{2}\ln\Bigl(\frac{P_{00} T}{2\tilde{\alpha}_R}\Bigr) \;=\; \frac{\mu\,\rho}{\lambda+\mu-1}\ln\left(\Bigl(1-\rho\frac{1-\lambda-2\mu}{1-\lambda-\mu}\Bigr)\frac{T}{
2\tilde{\alpha}_R}\right).
\end{eqnarray*}
Near criticality, where $P_{00} \approx 1$, both parts of the entropy production are proportional to the density of particles $\rho$, i.e.
\begin{equation}
\hrev/\rho \;\simeq\; \frac{\mu \ln (\lambda/\mu)}{\lambda+\mu-1} \,\qquad
\hirr/\rho \;\simeq\; \frac{\mu \ln (T/2\tilde{\alpha}_R)}{\lambda+\mu-1}\,.
\end{equation}
These quotients vary along the transition line, as shown in the right panel of Fig.~\ref{fig:cpent}. 

%======================================================
%======================================================

%================================================================================
\subsection{Growth process with roughening transition}
%================================================================================

\begin{figure}[t]
\centering\includegraphics[width=130mm]{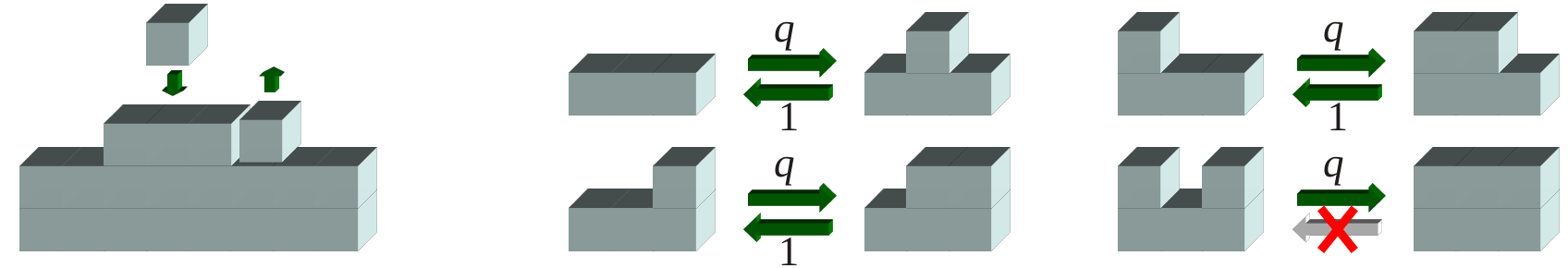}
\vspace{-2mm}
\caption{Dynamical rules of a model for interface growth which exhibits a depinning transition in 1+1 dimensions (see text). \label{fig:growthmodel}}
\end{figure}

\begin{figure}[t]
\centering\includegraphics[width=110mm]{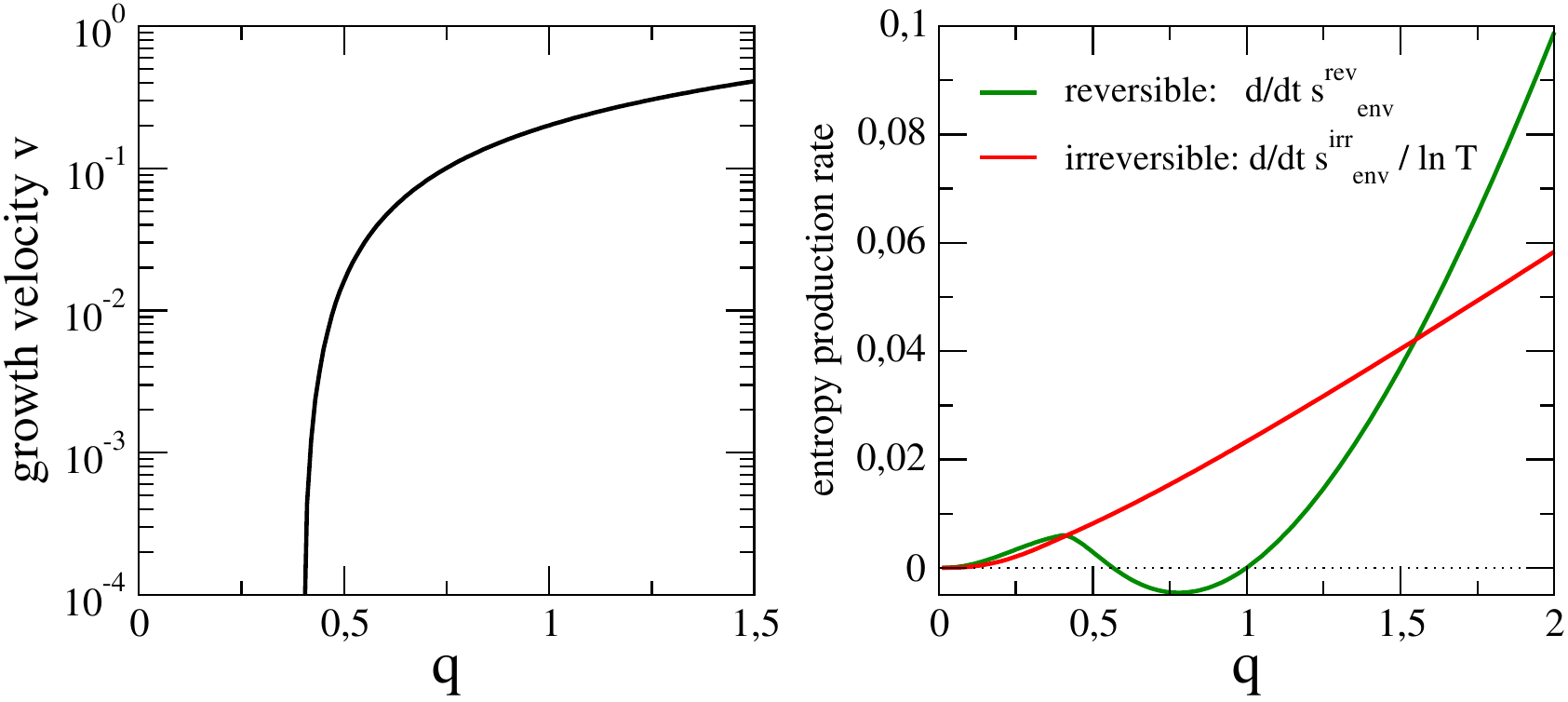}
\vspace{-2mm}
\caption{Model for interface growth. Left: Growth velocity as a function of the parameter $q$, indicating the unbinding transition at $q_c \approx 0.4$. Right: Reversible and irreversible entropy production produced by a system with $L=10^4$ sites monitored over $T=10^6$ Monte Carlo sweeps (see text). \label{fig:growth}}
\end{figure}

As another example, let us consider a model of a growing 1+1-dimensional interface which exhibits a roughening transition~\cite{Alon}. The configuration of the model is described by an interface in terms of a height variable $h_i\in\mathbb{Z}$ at the sites $i=1,\ldots,L$. In addition, it is assumed that the interface obeys the restricted solid-on-solid (RSOS) condition
\begin{equation}
|h_{i+1}-h_i| \leq 1.
\end{equation}
As illustrated in Fig.~\ref{fig:growthmodel}, the model evolves random-sequentially by deposition of particles with rate $q$ and spontaneous removal with rate $1$. As a special feature, evaporation from plateaus is forbidden, which means that one of the backward rates vanishes (the entropy production for a reversible non-zero rate was studied in \cite{Andre1,Andre2}). This vanishing rate, which is marked by a red cross in  Fig.~\ref{fig:growthmodel}, ensures that a layer, once completed, cannot be evaporated again, pinning the interface at a spontaneously selected height if the growth rate is very small. However, when the growth rate exceeds a certain critical threshold at $q_c \approx 0.4$ the interface unbinds from the bottom layer, leading to a roughening transition. In the moving phase the interface has a finite average velocity $v=\langle \dot h \rangle >0$, as shown in the left panel of Fig.~\ref{fig:growth}.

The first three processes shown in Fig.~\ref{fig:growthmodel} are reversible and therefore contribute with $\Delta S_{env} = \pm \ln q$ for deposition and evaporation, respectively. The last process, however, is irreversible and contributes with $\ln(p_f(q)qT/\tilde{\alpha}_R)$, where $p_f(q)$ denotes the probability to find three adjacent sites at the same height. These contributions are added up whenever the respective processes take place. 

The result is shown in the right panel of Fig.~\ref{fig:growth}. Interestingly, the reversible part of the entropy production exhibits a local maximum at the transition point, whereas the irreversible part of the entropy production is completely insensitive to the transition. 

%======================================================
%======================================================
%
%======================================================
%======================================================

\newpage

%==========================================================================